\documentclass{elsart5p}

\usepackage{graphicx}

\usepackage{amssymb}

\newcommand{\Vec}[1]{\mbox{\boldmath$#1$}}
\begin{document}

\begin{frontmatter}



\title{Unconventional superconductivity originating from\\ 
disconnected Fermi surfaces in the iron-based compound}


\author[HA]{Hideo Aoki}
\address[HA]{Department of Physics,  University of Tokyo, Tokyo 113-0033, Japan}

\begin{abstract}
The iron-based LaFeAsO$_{1-x}$F$_x$ 
recently discovered by Hosono's group is a fresh theoretical challenge as a new class of high-temperature superconductors.  
Here we describe the electronic structure 
of the material and the mechanism of superconductivity.   
We start with constructing a tight-binding model in terms of the maximally localized Wannier orbitals from a first-principles electronic structure calculation, which has turned out to involve all the five Fe 3d bands. This is used to calculate the spin and charge susceptibilities with the random phase approximation. The spin susceptibility has peaks around ${\Vec k} = (\pi, 0), (0, \pi)$ arising from a nesting across disconnected Fermi surface pockets.  We have then plugged the susceptibilities into the linearised Eliashberg equation.  For the doping concentration $x = 0.1$, we obtain an unconventional s-wave pairing, which is 
roughly an extended s in that the gap changes sign between the Fermi pockets, 
but the gap function is actually a 5$\times$5 matrix.  Its experimental implications are also discussed.
\end{abstract}

\end{frontmatter}

\section{Introduction}
\begin{figure}[h]
\begin{center}
\includegraphics[width=8.7cm,clip]{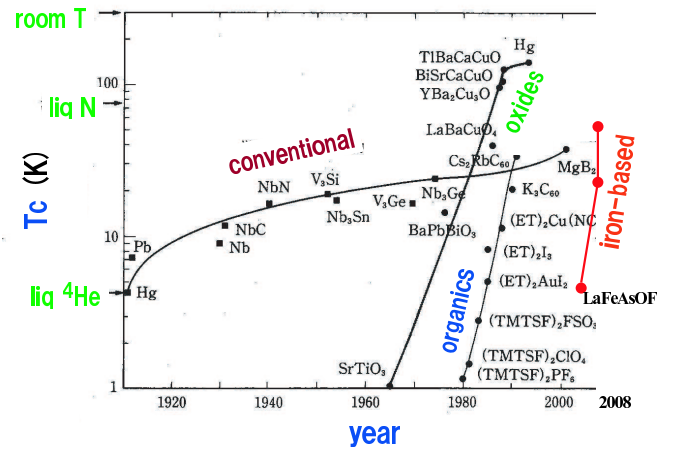}
\caption{(color online)
Tc against calendar year for various classes of superconductors.
\label{fig.Tc}}
\end{center}
\end{figure}

The discovery of the high-$T_C$ cuprate was epoch-making in that the material 
has opened a new avenue of the correlated electron systems 
and associated electron mechanism of superconductivity.  
However, we are still some way from a complete understanding, 
and a new class of materials has been desired, which 
would be a testbench for various ideas and concepts, 
both experimental and theoretical, in the 
electron correlation.  
So the discovery of superconductivity in 
the iron-based pnictide LaFeAsO doped with fluorine 
(LaFeAsO$_{1-x}$F$_x$) discovered by Hosono's group\cite{Hosono} is particularly 
welcome.  Indeed, this is the the first non-copper family 
of compounds whose 
$T_c$ exceeds 50 K in the subsequent studies (Fig.\ref{fig.Tc}).  

This immediately stimulates renewed interests in the 
electron mechanism of high $T_c$ superconductivity.  
Especially interesting, in our view, 
is that the way in which the correlation effects such as magnetism and superconductivity appear in strongly correlated 
electron systems is very sensitive to the underlying band structure.  We can envisage to 
exploit the ``fermiology in correlated electron systems" 
to explore the possibility of manipulating superconductivity 
by choosing the shape of the Fermi surface.\cite{fermiology}

In this paper we first briefly look at unconventional superconductivity 
as a background.  We then give a description of 
the electronic structure and a possible mechanism for 
superconductivity for the iron-based compound.  
Theoretically, the material turns out to have 
a curious multiband structure, with the Fermi surface 
comprising some pockets.  
In order to investigate the pairing mechanism, 
here we first construct an  
electronic model for the iron-based compound with 
the ``downfolding" based on a 
first-principles electronic structure calculation.  The minimal model 
turns out to contain all the five Fe $d$ orbitals. 
We then apply the five-band 
random-phase approximation (RPA) 
to solve the Eliashberg equation. 
We conclude that a nesting between multiple Fermi 
surface (pockets) results in a development of peculiar 
spin fluctuation modes, which in turn 
realises an unconventional 
$s$ (roughly extended $s$) pairing.\cite{1stpaper} 
The result is intriguing as one realisation of the 
idea of fermiology in general and the idea of 
the ``disconnected Fermi surface" in particular\cite{kuroki_disconnecteds}.
Finally we shall summarise that, as opposed the 
strongly correlated, half-filled, one-band cuprate as a doped Mott insulator, 
the iron compound is moderately correlated, dilute multi-band 
system.  An outlook for the iron compound is also given.

\section{Unconventional superconductivity}

After the discovery of the cuprate, a number of other classes of 
superconductors have been discovered, among which are 
the ruthenate, the cobaltate, the Hf compound and the 
Ce compound.  These have given a strong impetus towards 
the study of electron mechanism of superconductivity.  
In the conventional mechanism, superconductivity arises 
from the electron-electron attraction mediated by phonons, 
where $T_C \sim 10$ K is one order of magnitude smaller than 
the phonon energy $\omega_D \sim 100$ K.  
In the electron mechanism, superconductivity arises 
from the electron-electron repulsion, where 
spin and charge fluctuations mediate pairing 
with $T_C \sim 100$ K, two orders of magnitude smaller than 
the electronic energy $E_F \sim 10000$ K.  

The pairing from repulsive interactions itself is nothing strange, 
if we allow anisotropic pairing symmetries accompanied 
by anisotropic gap functions, $\Delta({\Vec k})$.  
The point is, if we look at the BCS gap equation, 
\begin{equation}
\Delta({\Vec k})=
-\sum_{{\Vec k}'}V({\Vec k},{\Vec k}')\frac{\Delta({\Vec k}')}{2E({\Vec k}')}
{\rm tanh}\left(\frac{E({\Vec k}')}{2k_B T}\right),
\label{lingapeq}
\end{equation}
where $V({\Vec k},{\Vec k}')$ is the pairing interaction, 
we can immediately recognise that 
a repulsion acts as an attraction when $\Delta({\Vec k})$ 
changes sign across the typical momentum transfer 
(usually dictated by the dominant spin fluctuation mode). 
This is what is happening in the cuprate with a $d$-wave pairing.  

The question, then, is 
``why is $T_C \sim \frac{1}{100}E_F$ in the electron mechanism is 
so low?", as most transparently displayed by 
Uemura's experimental plot for $T_C$ against $T_F$\cite{Uemuraplot}.  
Theoretically, there are good reasons why $T_C$ is low: 
(a) the effective attraction mediated by the fluctuation 
is much smaller than the bare (repulsive) interaction. 
(b) Quasi-particles are short-lived due to large 
self-energy corrections from the electron correlation, and 
(c) the prerequisite anisotropy in pairing with 
the gap-function nodes, which usually intersect the 
Fermi surface, suppresses $T_C$.   We can then think of 
how we can manipulate these factors.  
Several years ago Kuroki and Arita\cite{kuroki_disconnecteds} 
have proposed that we can 
overcome the difficulty (c) by considering {\it disconnected} 
Fermi surfaces, on which we can pierce the nodes 
in between the Fermi pockets.  Each pocket is then fully gapped, 
with opposite signs across the pockets.  
While the disconnected Fermi surfaces may first seem artificial, 
it has been recognised that they actually occur in 
real materials including TMTSF\cite{tmtsf} and Co compounds\cite{KKNCOO}.

As for the spatial dimensionality, 
Arita et al have shown that two-dimensional (layered) systems are 
more favourable than 3D ones 
for the spin-fluctuation mediated superconductivity 
for ordinary lattices\cite{2d3d}.  
This is because the fraction of the phase space over which 
the pairing interaction is appreciable is larger in 2D.  
The result, also obtained by Monthoux and Lonzarich\cite{monthoux}, 
agrees with the empirical fact that recently discovered 
superconductors (cuprates, Co compound, Hf compound, CeCoIn$_5$, etc) 
are mostly layer-structured.  The iron-based compound 
has turned out to be no exception.

\section{LaFeAsO}

\begin{figure}[h]
\begin{center}
\includegraphics[width=8.7cm,clip]{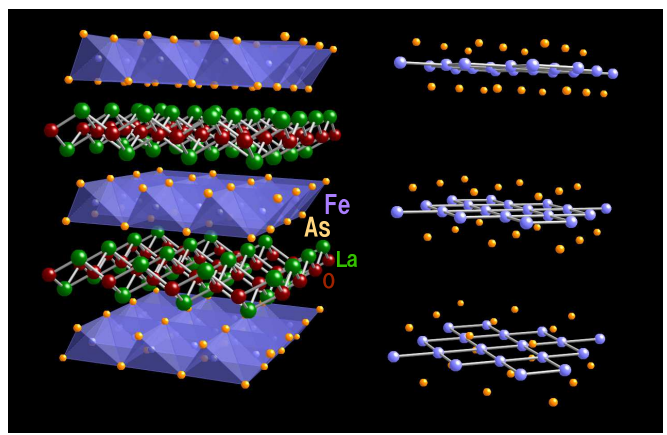}
\caption{(color online)
Crystal structure of the LaFeAsO.  The right panel 
picks up the FeAs layers.
\label{fig.structure}}
\end{center}
\end{figure}

LaFeAsO has a tetragonal layered structure (Fig.\ref{fig.structure}), 
where each Fe layer comprises edge-shared 
FeAs$_4$ tetrahedra with Fe atoms forming a 
square lattice.  Since the bands around $E_F$ have 
negligible LaO-layer components, we can concentrate on the Fe layer. 
The experimentally determined lattice constants 
are $a=4.03552$\AA and $c=8.7393$\AA, with 
two internal coordinates $z_{\rm La}=0.1415$ and $z_{\rm As}=0.6512$.
In the experimentally obtained phase diagram in 
the original discovery\cite{Hosono} and in an 
$\mu$SR study\cite{luetkens}arXiv0806.3533 
the (colinear; see below) SDW is taken over 
by the superconducting phase as the F-doping level is increased.

After the discovery of superconductivity, a surge of intensive 
studies on the material has ensued.  Since it is impossible to 
survey them here, let us here mention only a few examples with an extremely 
partial list of references.  
In the iron-based family, various compounds have been shown to 
exhibit superconductivity, among which are 
NdFeAsO$_{1-x}$F$_x$\cite{Ren08034234}, 
PrFeAsO$_{1-x}$F$_x$\cite{Ren08034283}, 
GdFeAsO$_{1-y}$\cite{Gd}, 
SmFeAsO$_{1-x}$F$_x$\cite{Ren08042053}, 
with $T_C$ now exceeding 56 K.  
Pressure effects have been studied for 
LaFeAsO$_{1-x}$F$_x$\cite{takahashi}, 
NdFeAsO$_{1-y}$\cite{takeshita}, etc. 
Superconductivity has also been found in 
iron-based materials with different layered structures 
that include BaFe$_2$As$_2$\cite{BaFe2As2} 
and FeSe\cite{FeSe}.  

We can first itemise the experimental results that indicate 
unconventional SC:

\begin{enumerate}
\item 
NMR, where absence of coherence peak, $1/T_1 \sim 1/T^3$, etc have 
been observed.\cite{nmr}

\item 
Electronic specific heat $\gamma \propto \sqrt{H}$, etc has 
been observed.\cite{specificheat}

\item 
Point-contact tunnelling, where zero-bias peak, 
indicative of a sign change across Fermi pockets, etc have 
been observed.\cite{pointcontact}

\item 
$\mu$SR is used to measure the superfluid density.\cite{mSR} 
Especially interesting is the Uemura plot, in which the iron compound 
is shown to reside more or less on the Uemura line, so 
in this sense the iron compound is at least as good as the cuprate.  

\item 
Raman spectra indicate a weak electron-phonon coupling.\cite{raman}

\item 
$H_{c2}, H_{c1}$ indicate a two-gap behaviour.\cite{Hc2} 

\item 
Photoemission spectroscopy,\cite{PES} 
and more recently ARPES depict the Fermi surface, gap, etc\cite{ARPES}.
      
\end{enumerate}

Experimental results for magnetic properties, especially interesting 
in relation to the mechanism of superconductivity, include the following:

\begin{enumerate}
\item 
M\"{o}ssbauer studies indicate a magnetic transition temperature 
$T_N \sim 140$ K with a magnetic moment 
$M = 0.25-0.35 \mu_B$ (at $T=4$ K)\cite{mossbauerKitao,klass}. 

\item 
Neutron diffraction studies indicate a colinear (stripe) 
SDW (see Fig.\ref{fig.chis}) in Fe moments with $T_N \sim 134$ K, 
$M \sim 0.35 \mu_B$ (at $T=8$ K)\cite{neutron}. 

\item 
$\mu$SR suggests a spontaneous muon spin precession below $T_N = 
138$ K\cite{muSRspontaneous}.
     
\item 
NMR shows $T_N = 142$ K\cite{nmrTN}. 

\item 
Transport measurements indicate an anomaly around $T = 
140-150$ K\cite{transport}.

\item 
X-ray diffraction shows a structural phase transition (tetragonal 
$\rightarrow$ orthorhombic) at $\simeq 150$ K\cite{xray}, 
which is close to, but somewhat deviates from, the magnetic 
transition temperature.

\end{enumerate}

Theoretically, if we first 
look at the periodic table, copper, in its Cu$^{2+}$ 
state, has a $3d^9$ electron configuration.  This makes the cuprate 
one-band, nearly half-filled system, where we have a spin 1/2 
at each site with strong antiferromagnetic fluctuations.  
By contrast, Fe, located in the middle of the table, 
has a $3d^6$ configuration, so multiple $d$ bands tend to be 
involved.  A point of interest is that 
we usually associate iron with ferromagnetism 
rather than with superconductivity\cite{ironSC}, 
and how the multiband structure for the iron compound 
gives rise to superconductivity. 

We then definitely require the electronic band structure to 
begin with.  
\begin{itemize}
\item Several groups have obtained  the band structure\cite{lebegue,Singh,ishibashi}. 
\item Theoretical suggestiongs on 
magnetic properties include an LDA+DMFT study 
which indicates no local moment for the Hund's 
coupling $J_H < 0.35$ eV\cite{haule}. 
\item An {\it ab initio} (cRPA) calculation 
estimats $J_H \simeq 0.2$ eV.\cite{NakamuraArita} 
\item An itinerant charachter of the compound 
suggested from several papers\cite{itinerant}.
\item  Sensitivity of the electronic structure against 
        the Fe-As distance has been suggested in 
        a maximally localised Wannier study that incorporates As $4p$ 
        orbitals\cite{Vildosola}.
\end{itemize}

\section{Model construction}

First two papers that study the mechanism of superconductivity are 
by Mazin et al\cite{Mazin} and by Kuroki et al\cite{1stpaper}.  
The former starts from the Fermi surface and construct a 
an effective RPA treatment for the spin susceptibility 
$\chi_0({\Vec q})/[1-J({\Vec q})\chi_0({\Vec q})]$ 
to predict the colinear SDW and 
an extended s-wave pairing.   
The latter starts with a model building via downfolding 
to actually solve Eliashberg's equation for the full five-band 
model.  Here we briefly describe this approach.

\begin{figure}[h]
\begin{center}
\includegraphics[width=8.7cm,clip]{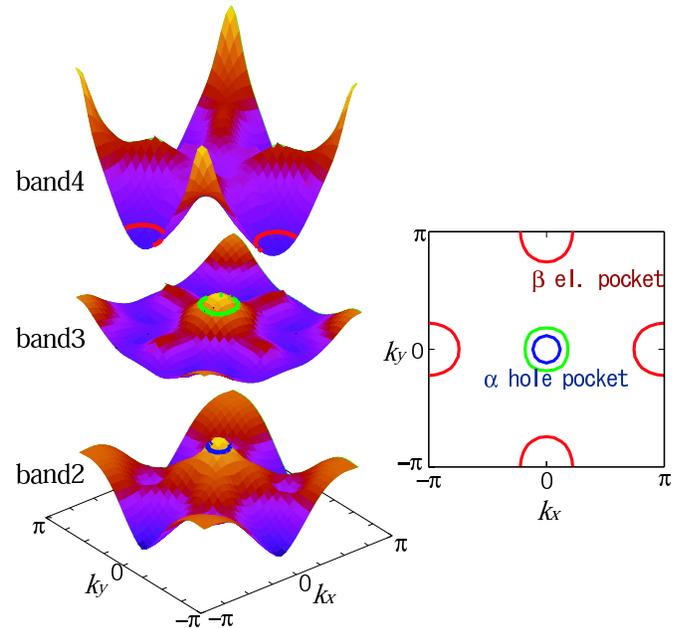}
\caption{(color online)
Left panel: 
The dispersion of the three bands that intersect $E_F$ 
in LaFeAsO$_{1-x}$F$_x$  with 10 \% doping ($x=0.1$). 
Right panel: Fermi surface (with the 
inter-layer hopping ignored).
\label{fig.dispersion}}
\end{center}
\end{figure}

\begin{figure}[h]
\begin{center}
\includegraphics[width=8.7cm,clip]{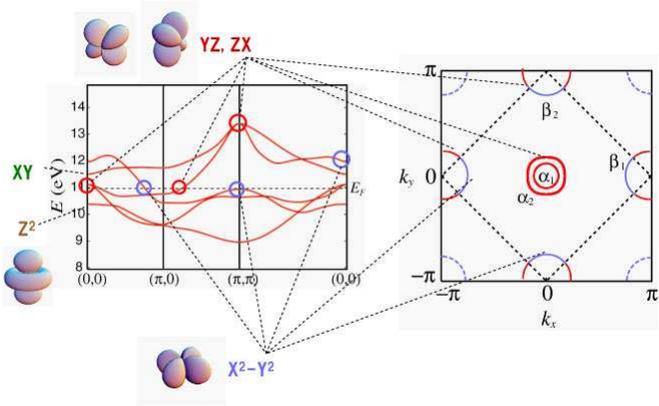}
\caption{(color online)
The band structure of the five-band model 
in the unfolded BZ, where the 
inter-layer hoppings are included.  
Orbital character is indicated for the 
bands and for the Fermi surface (right panel, 
with the original (dashed lines) and 
the unfolded (solid) BZ indicated).
Bottom panels depict (two equivalent 
realisations of) the colinear SDW. 
\label{fig.character}}
\end{center}
\end{figure}

The approach starts from a band structure 
(Figs.\ref{fig.dispersion}, \ref{fig.character}) 
with the 
density-functional approximation with plane-wave basis\cite{pwscf} 
for constructing the model.  
For that, we have employed a standard procedure with 
maximally localized 
Wannier functions (MLWFs)\cite{MaxLoc}. 
These MLWFs, centred at the two Fe sites in the unit cell, 
have five orbital symmetries (orbital 1: $d_{3Z^2-R^2}$, 
2: $d_{XZ}$, 3: $d_{YZ}$, 4: $d_{X^2-Y^2}$,
5: $d_{XY}$). 
Due to the tetrahedral coordination of As, 
there are two Fe atoms per unit cell, 
and $X, Y, Z$ refer to those in the original unit cell.  
The two Wannier orbitals in 
each unit cell are equivalent in that each Fe atom has the same 
local arrangement of other atoms.
We can thus take a unit cell that 
contains only one orbital per symmetry by 
unfolding the Brillouin zone (BZ),  
and we end up with an effective five-band model on a 
square lattice, where $x$ and $y$ axes are rotated by 
45 degrees from $X$-$Y$, 
to which we refer for all the wave vectors hereafter. 
We define the band filling $n$ as the number of electrons/number of sites
(e.g., $n=10$ for full filling). 
The doping level $x$ 
in LaFeAsO$_{1-x}$F$_x$ is related to the band filling as $n=6+x$.

The five bands are heavily 
entangled as shown in Fig.\ref{fig.character}, 
which reflects hybridisation 
of the five $3d$ orbitals due to the tetrahedral 
coordination of As $4p$ orbitals around Fe.
Hence we conclude that the minimal electronic model 
requires all the five bands.\cite{reduce} 
In Fig.\ref{fig.character}, 
the Fermi surface for $n=6.1$ (corresponding to $x=0.1$) 
obtained by ignoring the inter-layer hoppings 
is shown in the two-dimensional unfolded BZ.  
The shape of the Fermi surface is now being experimentally 
detected with ARPES.\cite{ARPES}

The Fermi surface consists of 
four pieces (pockets in 2D when the 
inter-layer hoppings are neglected):   
two concentric hole pockets (denoted as $\alpha_1$, $\alpha_2$) 
centred around $(k_x, k_y)=(0,0)$, two electron pockets 
around $(\pi,0)$ $(\beta_1)$ or $(0,\pi)$ $(\beta_2)$, respectively. 
$\alpha_i$ ($\beta_i$) corresponds to the 
Fermi surface around the $\Gamma$Z (MA) line (in the original BZ) 
in Singh's band calculation.\cite{Singh}
In addition, we notice that a portion of the band 
near $(\pi,\pi)$ is flat and 
touches the $E_F$ at $n=6.1$, so that 
the portion acts as a ``quasi Fermi surface $(\gamma)$'' around $(\pi,\pi)$. 
The orbital character is again entangled: $\alpha$ and portions of $\beta$ near 
BZ edge has mainly $d_{XZ}$ and $d_{YZ}$ character, while the portions of 
$\beta$ away from the BZ edge and $\gamma$ have 
mainly $d_{X^2-Y^2}$ orbital character.

\section{Five-band RPA results for magnetism and pairing}
Having constructed the model, we move on to the RPA calculation 
in the 2D model.  For the many-body part of the Hamiltonian,  
we consider the standard interaction terms that comprise 
the intra-orbital Coulomb $U$, the inter-orbital 
Coulomb $U'$, the Hund's coupling $J$, and the pair-hopping $J'$. 
The modification of the band structure due to 
the self-energy correction is not taken into account. 
For the five-band model, 
Green's function is a $5\times 5$ matrix in the orbital representation, while 
each of the spin and orbital susceptibilities $\chi_{l1,l2,l3,l4} 
(l_i = 1,...,5)$ is a $25\times 25$ matrix.  
The Green's function and 
the effective pairing interactions, obtained from the susceptibilities,  
are plugged into the linearised Eliashberg equation, and 
the gap function as a $5\times 5$ matrix is obtained 
along with the eigenvalue $\lambda$. 
$32\times32$  ${\Vec k}$-point meshes and 1024 Matsubara frequencies are taken. 
We find that the spin fluctuations dominate over orbital fluctuations as far as 
$U>U'$, so we can characterise the system with the spin susceptibility.
We denote the largest eigenvalue of the spin 
susceptibility matrix for $i \omega_n=0$ 
as $\chi_s(\Vec{k})$. 
The gap function matrix at the lowest Matsubara frequency 
is transformed into the band representation by 
a unitary transformation, and its 
diagonal element for band $i$ is denoted as 
$\phi_i(\Vec{k})$.

\begin{figure}
\begin{center}
\includegraphics[width=7.0cm,clip]{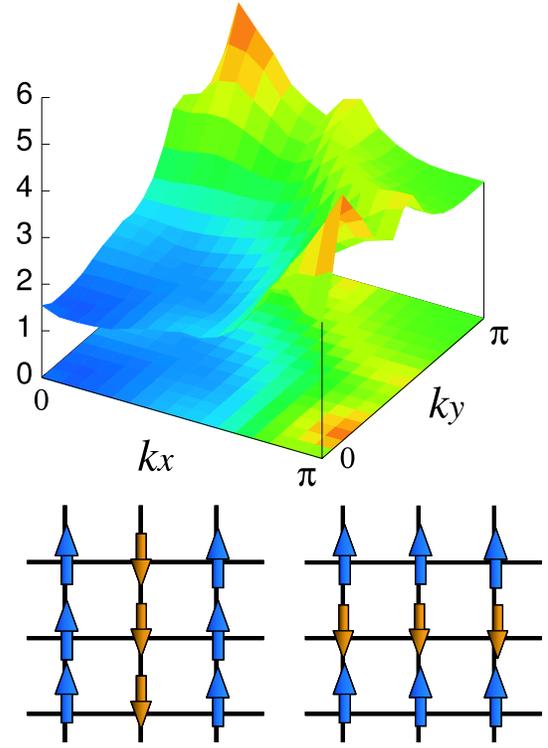}
\caption{(color online)
5-band RPA result for the spin susceptibility 
for 10\% doped LaFeAsO$_{1-x}$F$_x$ for 
$U=1.2$, $U'=0.9$, $J=J'=0.15$, and $T=0.02$ (in eV).  
Bottom panels depict the colinear SDW for 
$\Vec{k}=(\pi,0), (0,\pi)$.
\label{fig.chis}}
\end{center}
\end{figure}

\begin{figure}
\begin{center}
\includegraphics[width=9.0cm,clip]{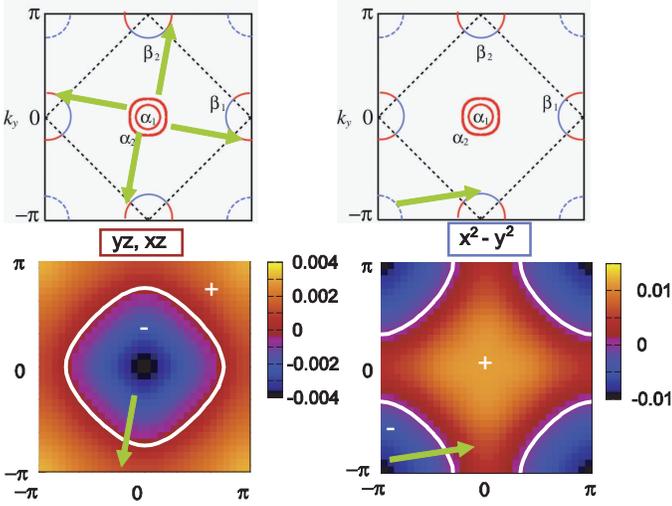}
\caption{(color online)
Bottom panels: 
Diagonal elements of the gap function matrix in the orbital 
representation for the $XZ, YZ$ (left) and $X^2-Y^2$ (right) 
in the 5-band RPA for 10\% doped LaFeAsO$_{1-x}$F$_x$ 
for the same parameter set as in the previous figure.  
Nesting vectors are indicated in the top panels.
\label{fig.gap}}
\end{center}
\end{figure}

Let us look at the result for the spin susceptibility 
$\chi_s$ for $U=1.2$, $U'=0.9$, $J=J'=0.15$, and $T=0.02$ 
(all in units of eV) in Fig.\ref{fig.chis}. 
The spin susceptibility has peaks around 
$(k_x, k_y) = (\pi,0)$, $(0,\pi)$ (i.e., a colinear SDW), which 
reflects the Fermi surface nesting around $\sim (\pi,0),(0,\pi)$ 
across $\alpha$ and $\beta$ 
and also $\beta$ and $\gamma$ in Fig.\ref{fig.character}. The 
spin structure is 
consistent with the neutron scattering experiments 
in which the colinear structure is observed 
at low temperatures for the undoped case.\cite{neutron}

\begin{figure}
\begin{center}
\includegraphics[width=8.7cm,clip]{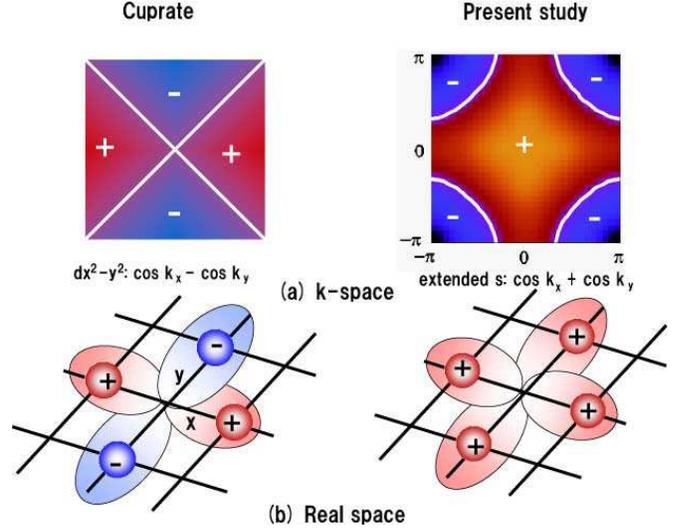}
\caption{(color online)
Gap function symmetry is schematically shown for 
the cuprate (left panels) and the present result for 
the iron compound (right), in the $k$-space (top panels) 
and in the real space (bottom).  
\label{fig.gaprealspace}}
\end{center}
\end{figure}

For superconductivity, we show in Fig.\ref{fig.gap} 
the diagonal elements of the gap function matrix 
for orbitals $d_{YZ}$, $d_{ZX}$, and $d_{X^2-Y^2}$.
The gap is a curious extended $s$-wave,\cite{Mazin,1stpaper} where 
the gap changes sign across the $\alpha-\beta$ or $\beta-\gamma$ 
nesting vector $\sim (\pi,0),(0,\pi)$ at which the spin fluctuations 
develop.  
The sign change 
is analogous to those in models studied by Bulut {\it et al.},\cite{Bulut} 
and Kuroki and Arita\cite{kuroki_disconnecteds}. 
To be more precise for the multiband system, 
the magnitude of gap of the $d_{X^2-Y^2}$ orbital 
turns out to be large compared to other orbitals, 
which indicates that the $d_{X^2-Y^2}$ orbitals play the 
main role in the superconductivity.
The extended s ($\sim {\rm cos}k_x + {\rm cos}k_y$ in ${\Vec k}$ space) 
should look in real space as depicted in Fig.\ref{fig.gaprealspace}.

\begin{figure}
\begin{center}
\includegraphics[width=5.5cm,clip]{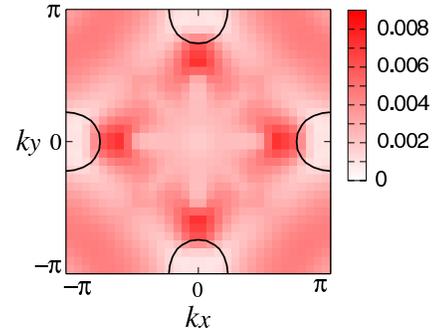}
\caption{(color online)
5-band RPA result for the gap function, 
$\sqrt{(\hat{\phi}\hat{\phi}^\dagger)_{44}}$ 
for the same parameter set as in the previous figures. 
\label{fig.Delta44}}
\end{center}
\end{figure}

However, we have to be careful in analysing the gap, since we have a 
multiband system at hand, where the band character changes along the 
Fermi surface (Fig.\ref{fig.character}).  
When we concentrate on the main component ($d_{X^2-Y^2}$ orbitals) 
of the gap function, the magnitude of the gap varies 
along the $\beta$ Fermi surface, which is because the 
character of the Fermi surface changes, 
as we approach the BZ boundary, to $d_{YZ}$, $d_{ZX}$ 
for which the gap is smaller.  More importantly, 
we should always be aware that the gap function is 
a 5$\times$5 matrix, where off-diagonal elements 
are rather significant due to the heavy entanglement of the bands.  
This should have experimental implications for various quantities.  
An importance of matrix character 
of the gap is illustrated in Fig.\ref{fig.Delta44} that displays 
the quantity $(\hat{\phi}\hat{\phi}^\dagger)_{44}$, where $\hat{\phi}$  is the 
gap matrix and 44 denotes the diagonal element of band 4. 
As shown in the figure, this quantity 
is fully gapped over the entire BZ.  A remnant of the nodal lines of the 
diagonal element appears as a dip that intersects the $\beta$ Fermi surface, 
which may look like a node e.g. at not low enough temperatures. 
The degree of the variation of the gap 
may be determined experimentally from e.g. tunnelling spectroscopy or 
ARPES.  Also interesting is how the penetration depth, 
$\lambda \sim n_s/m^*$ with $n_s$ being the superfluid 
density in the one-band system, should be given in the multiband 
system.

\section{Conclusion}
To summarise, we have very briefly reviewed the iron compound, and 
then described a theory in which a five-band electronic 
model is constructed as a minimum microscopic model. 
Applying a five-band RPA to this model,  
we have found that spin fluctuation modes around $(\pi,0),(0,\pi)$ 
develop due to the nesting between disconnected Fermi surfaces.
Based on the linearised Eliashberg equation, 
we have concluded that spin fluctuation modes 
realise an unconventional, extended $s$-wave pairing, where 
the diagonal elements of the gap matrix change sign across the 
nesting vector. The $d_{X^2-Y^2}$ orbital is the main 
component in the superconductivity gap function, although 
the gap function as a $5\times 5$ matrix may be important.  

Overall, while the cuprate is strongly correlated ($U>W$ 
with $W$ the band width), 
one-band and nearly half-filled system, the iron compound is 
moderately correlated ($U\sim W$)\cite{NakamuraArita}, 
multi-band and dilutely filled system. 
This poses a challenging problem of 
whether the iron compound can exceed the cuprate.  
Relevance to experimental results, especially 
$\mu$SR including the Uemura plot, is also a 
pressing future problem.    

The work described here is a collaboration with 
Kazuhiko Kuroki, Seiichiro Onari, 
Ryotaro Arita, Hidetomo Usui, Yukio Tanaka 
and Hiroshi Kontani.  
Numerical calculations were performed at the facilities of
the Information Technology Center, University of Tokyo, 
and also at the Supercomputer Center,
ISSP, University of Tokyo. 
This study has been supported by 
Grants-in-Aid for Scientific Research from  MEXT of Japan and from 
the Japan Society for the Promotion of Science.

%

\end{document}